\RequirePackage{fix-cm}
\documentclass[smallextended]{svjour3}       
\smartqed  
\usepackage{amssymb}
\usepackage{amsmath}
\usepackage{color}
\usepackage{graphicx}
\usepackage[justification=centering]{caption}
\usepackage{multirow}
\usepackage{extarrows}
\usepackage{booktabs}
\usepackage{bigstrut}
\usepackage{cite}
\begin{document}

\title{Quantum Privacy-Preserving Price E-Negotiation
}


\author{Wen-Jie Liu     \and
        Chun-Tang Li         \and
        Yu Zheng     \and
        Yong Xu     \and
        Yin-Song Xu  
}


\institute{
           W.-J. Liu \and
           Y. Zheng
           \at Jiangsu Engineering Center of Network Monitoring, Nanjing University of Information Science \& Technology, Nanjing 210044, P.R.China
           \\
           \email{wenjiel@163.com}
           \and
           W.-J. Liu   \and
           C.-T Li     \and
           Y. Zheng   \and
           Y. Xu     \and
           Y.-S. Xu
           \at School of Computer and Software, Nanjing University of Information Science \& Technology, Nanjing 210044, P.R.China
 }

\date{Received:  / Accepted: date}

\maketitle

\begin{abstract}
Privacy-preserving price e-negotiation (3PEN) is an important topic of secure multi-party computation (SMC) in the electronic commerce field, and the key point of its security is to guarantee the privacy of seller's and buyer's prices. In this study, a novel and efficient quantum solution to the 3PEN problem is proposed, where the oracle operation and the qubit comparator are utilized to obtain the comparative results of  buyer's and seller's prices, and then quantum counting is executed to summarize the total number of products which meets the trading conditions. Analysis shows that our solution not only guarantees the correctness and the privacy of 3PEN, but also has lower communication complexity than those classical ones.

\end{abstract}

\section{Introduction}
\label{intro1}
Negotiation is a dialogue between two or more people or parties intended to reach a beneficial outcome over one or more issues where a conflict exists with respect to at least one of these issues. This beneficial outcome can be for all of the parties involved, or just for one or some of them. Thus, negotiation plays an important role in the electronic commerce, artificial intelligence, and even political fields. As we know, electronic trading(e-trading) has become increasingly pervasive in the business world. And electronic negotiation (e-negotiation) is a critical process of e-trading where the negotiating parties with different preferences, criteria, and constraints try to maximize their gains and reach a consensus on the terms of trading such as prices, delivery dates, and payment terms~\cite{CA15,AM12}. E-negotiation is motivated by its potential to provide business partners with more efficient processes, enabling them to arrive at more satisfying agreements in less time~\cite{AN12,CA09}.

In addition to efficiency improvements, the security of the exchanged data  in e-negotiation has also been concerned~\cite{LA00,HA02,YM05}. Limthanmaphon \emph{et al}.~\cite{LA00} proposed a agent-based negotiation model based on the digital signature,  which provides the concurrency control and the transactions security in e-negotiation. He and Leung~\cite{HA02} proposed three approaches for enhancing trust in e-transaction including: enforcement of third parties, chunking e-negotiation transactions into smaller chunks, and enforcement of social laws. Yang~\cite{YM05} proposed a secure protocol to protect data the agent collects and agent itinerary based on onion routing scheme. These protocols ensure integrity of agent itinerary, anonymity of non-neighbor hosts in the agent itinerary, and fault-tolerance of remote hosts. However, they do not address the privacy of participants.

Secure multi-party computation(SMC)~\cite{YP82} is a particular kind of cryptography with the goal of creating methods for parties to jointly compute a function over their inputs while keeping those private inputs in secret. As a practical topic of SMC, privacy-preserving price e-negotiation (3PEN) aims to guarantee the seller's and the buyer's privacy in electronic commerce. Based on distributed SMC, Chakraborty \emph{et~al}.~\cite{CP05} first proposed two privacy-preserving negotiation protocols (C05): one is non-discriminatory and the other is discriminatory, where the price and the demands of the buyers are not disclosed to anyone before entering the final agreement. Al-Jaljouli \emph{et~al}.~\cite{AS07} proposed a mobile agent-based e-negotiation protocol (A07) which protects the security properties of privacy, authentication, non-repudiation, anonymity, and integrity. It sholud be noted that these protocols have communication complexity $O(N)$, here $N$ is the numbers of products.

With the advent of quantum computation, classical cryptosystems, including the symmetric AES cryptosystem and the asymmetric RSA cryptosystem, are facing enormous threatens and challenges from quantum computer. Fortunately, quantum cryptography can provide the unconditional security, which is guaranteed by some physical principles of quantum mechanics, to resist against its impact. On the other hand, quantum parallelism makes it possible to greatly speed up solving some specific computational tasks, such as quantum factoring~\cite{SA94} and quantum searching~\cite{GA96}. With quantum mechanics utilized in the information processing, many important research findings are presented in recent decades, such as quantum key distribution (QKD)~\cite{BQ84,EQ91,LM12}, quantum key agreement (QKA)~\cite{CQ10,HE17,LM17,LA18}, quantum secure direct communication (QSDC)~\cite{LWJ09,LZH17,LZH18,ZJ18}, quantum private comparison (QPC)~\cite{YA09,LQ13,LS14,LS014,LC14}, quantum sealed-bid auction (QSBA)~\cite{NS09,LA14,LM16}, quantum state remote preparation (QRSP)~\cite{LWJ15,TX18,MM18}, quantum private query (QPQ)~\cite{GF12,LA19} and even quantum machine learning algorithms~\cite{LQ013,LQ18,LA019}. These above findings have shown the potential power in either the efficiency improvements or the security enhancements.

In this study, a quantum privacy-preserving price e-negotiation (Q3PEN) protocol,  i.e., an efficient quantum solution to the 3PEN problem, is proposed. For the convenience of description, we take the two-party negotiation as an example, and it can be easily extended to the multi-party scenario. In this protocol, the oracle operations are performed to prepare the  corresponding state, the qubit comparator is applied to obtain the final state which represents the comparative results of the two-party's prices, and finally quantum counting is executed to summarize the number of products which meets the trading conditions (i.e., the numbers of those the buyer's bid is higher than the seller's selling price surpass a certain threshold). The purpose of our work is to reduce the communication complexity and protect the two-party's privacy.

The rest of this paper is organized as follows. Some definitions about Q3PEN are given  in the next section, and a novel and efficient Q3PEN protocol is presented in Sect. 3. The correctness, privacy and efficiency analysis of our protocol are discussed in Sect. 4, and then Sect. 5 concludes the paper and makes some further discussions.

\section{Definitions of Q3PEN}
\label{sec:2}

Before introducing the procedure of Q3PEN protocol, we firstly make some definitions. Without loss of generality, we suppose there are two parties, the buyer Alice and the seller Bob, and the formal definitions of Q3PEN problem and Q3PEN protocol are given as below.

\textbf{Definition 1 (Q3PEN problem)} Bob has $N$ products with the minimum selling prices $B$=$(b_1^{},b_2^{}, \cdots, b_N^{})$, and Alice wants to buy these products under the maximum buying prices  $A$=$(a_1^{},a_2^{}, \cdots, a_N^{})$.  Then Alice and Bob jointly judge whether the number of the products meets the trading condition  $\sum\limits_{i = 1}^N {f({a_i},{b_i})}  \ge \varepsilon $  without leaking their private prices, where $f({a_i},{b_i}) \in \{ 0,1\} $ is the comparison function $f({a_i},{b_i}) = ({a_i} \ge {b_i})$,  and $\varepsilon $  is a predefined threshold (in general, $1 \ll \varepsilon  \le N$).  If the condition is satisfied, the two parties trade, otherwise they cancel the negotiation.

\textbf{Definition 2 (Q3PEN protocol)} Alice and Bob input their buying and selling price set  $A$, $B$, respectively. After executing this protocol, they can obtain the result of  $\sum\limits_{i = 1}^N {f({a_i},{b_i})}  \ge \varepsilon $, and decide whether they trade or not. In addition, the protocol should satisfy three properties as below,

\textbf{Correctness:} Two honest parties Alice and Bob can get the correct result of  $\sum\limits_{i = 1}^N {f({a_i},{b_i})}  \ge \varepsilon $.

\textbf{Alice's Privacy:} Bob cannot learn any secret information about Alice's prices $A$ except the possible information deduced from the final result of  $\sum\limits_{i = 1}^N {f({a_i},{b_i})}  \ge \varepsilon $  and his prices $B$.

\textbf{Bob's Privacy:} Alice cannot get any secret information about Bob's prices $B$ except the possible information deduced from $\sum\limits_{i = 1}^N {f({a_i},{b_i})}  \ge \varepsilon $  and  $A$.

\section{Protocol of Q3PEN}
\label{sec:3}
Referring to Definition 2, we propose a six-step Q3PEN protocol as below (shown in Fig. 1).
\begin{figure}[htb]
\centering
  \includegraphics[width=3.4in]{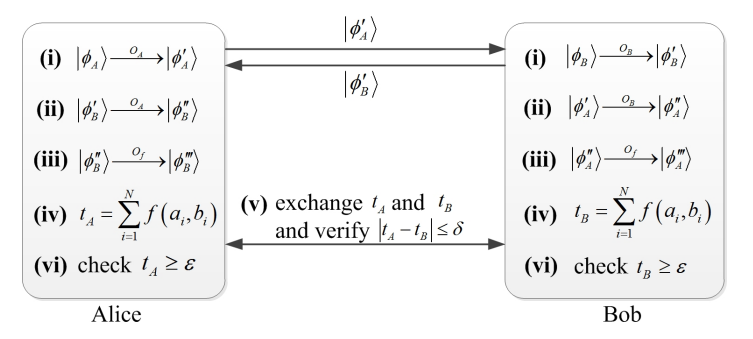}
  \DeclareGraphicsExtensions.
 \caption{The six-step procedure of Q3PEN protocol.}
\label{Fig. 1 }
\end{figure}

Step 1. Alice and Bob respectively prepare their initial states $\left| {{\phi _A}} \right\rangle$ and $\left| {{\phi _B}} \right\rangle$ which are both in state $\frac{1}{{\sqrt N }}\sum\limits_{i = 1}^{N} {\left| i \right\rangle  \otimes {{\left| 0 \right\rangle }^{ \otimes d}}}$, where $d = \left\lceil {\log \left( {\mathop {\max }\limits_{i,{\kern 1pt} {\kern 1pt} k} ({a_j},{b_k}) + 1} \right)} \right\rceil$ ($1 \le j,k \le N$). Then Alice (Bob) applies the oracle operation ${O_A}$ (${O_B}$) on  $\left| {{\phi _A}} \right\rangle$ ($\left| {{\phi _B}} \right\rangle$) (shown in Fig. 2).
\begin{equation}\label{1}
\frac{1}{{\sqrt N }}\sum\limits_{i = 1}^N {\left| i \right\rangle  \otimes {{\left| 0 \right\rangle }^{ \otimes d}}} \xrightarrow{{{O_A}}} \frac{1}{{\sqrt N }}\sum\limits_{i = 1}^N {\left| i \right\rangle \left| {{a_i}} \right\rangle }
\end{equation}
\begin{equation}\label{2}
\frac{1}{{\sqrt N }}\sum\limits_{i = 1}^{N} {\left| i \right\rangle  \otimes {{\left| 0 \right\rangle }^{ \otimes d}}} \xrightarrow{O_B} \frac{1}{{\sqrt N }}\sum\limits_{i = 1}^{N} {\left| i \right\rangle \left| {{b_i}} \right\rangle}
\end{equation}
Then Alice (Bob) sends the result state $\left| {{{\phi '}_A}} \right\rangle  = \frac{1}{{\sqrt N }}\sum\limits_{i = 1}^{N} {\left| i \right\rangle \left| {{a_i}} \right\rangle } $ ($\left| {{{\phi '}_B}} \right\rangle  = \frac{1}{{\sqrt N }}\sum\limits_{i = 1}^{N} {\left| i \right\rangle \left| {{b_i}} \right\rangle }$)  to Bob (Alice).

\begin{figure}[htb]
\centering
  \includegraphics[width=2.8in]{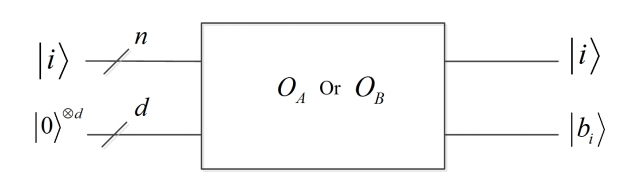}
  \DeclareGraphicsExtensions.
 \caption{The oracle operation Alice or Bob applied on the initial state $ \frac{1}{{\sqrt N }}\sum\limits_{i = 1}^{N} {\left| i \right\rangle  \otimes {{\left| 0 \right\rangle }^{ \otimes d}}}$. Here $n = \left\lceil {\log (N + 1)} \right\rceil$. }
\label{Fig. 2 }
\end{figure}

Step 2. After receiving $\left| {{{\phi '}_A}} \right\rangle$ ($\left| {{{\phi '}_B}} \right\rangle$)  from Alice (Bob), Bob (Alice) further performs the oracle operation ${O_B}$ (${O_A}$) between  $\frac{1}{{\sqrt N }}\sum\limits_{i = 1}^N {\left| {i} \right\rangle }$ in $\left| {{{\phi '}_A}} \right\rangle$ ($\left| {{{\phi '}_B}} \right\rangle$) and the auxiliary state ${\left| 0 \right\rangle ^{ \otimes d}}$, which is similar to the oracle operation in Step 1.
\begin{equation}\label{3}
\frac{1}{{\sqrt N }}\sum\limits_{i = 1}^{N} {\left| i\right\rangle \left| {{a_i}} \right\rangle \otimes {{\left| 0 \right\rangle }^{ \otimes d}}} \xrightarrow{{{O_B}}}\frac{1}{{\sqrt N }}\sum\limits_{i = 1}^{N} {\left| i \right\rangle \left| {{a_i}} \right\rangle \left| {{b_i}} \right\rangle }
\end{equation}
\begin{equation}\label{4}
\frac{1}{{\sqrt N }}\sum\limits_{i = 1}^{N} {\left| i\right\rangle \left| {{b_i}} \right\rangle  \otimes {{\left| 0 \right\rangle }^{ \otimes d}}} \xrightarrow{{{O_A}}}\frac{1}{{\sqrt N }}\sum\limits_{i = 1}^{N} {\left| i\right\rangle \left| {{b_i}} \right\rangle \left| {{a_i}} \right\rangle }
\end{equation}
Then Bob (Alice) obtains a new state  $\left| {{{\phi ''}_A}} \right\rangle  = \frac{1}{{\sqrt N }}\sum\limits_{i = 1}^{N} {\left| i\right\rangle \left| {{a_i}} \right\rangle \left| {{b_i}} \right\rangle }$ ($\left| {{{\phi ''}_B}} \right\rangle  = \frac{1}{{\sqrt N }}\sum\limits_{i = 1}^{N} {\left| i\right\rangle \left| {{b_i}} \right\rangle \left| {{a_i}} \right\rangle }$).

Step 3. Bob (Alice) applies another oracle operation ${O_f}$  on $\left| {{{\phi ''}_A}} \right\rangle$ ($\left| {{{\phi ''}_B}} \right\rangle $) and an auxiliary qubit $\left| 0 \right\rangle$ (shown in Fig. 3),
\begin{figure}[htb]
\centering
  \includegraphics[width=3.2in]{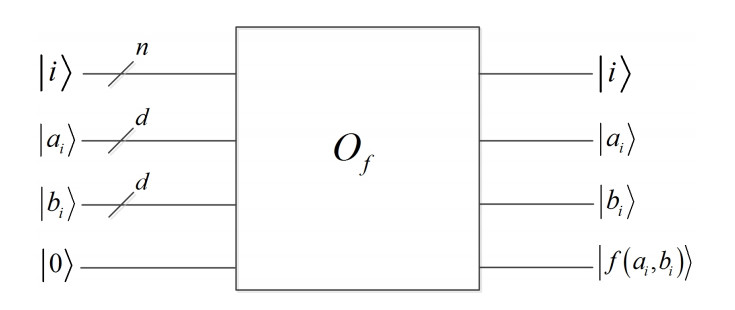}
  \DeclareGraphicsExtensions.
 \caption{The circuit of the oracle operation applied on state $ \frac{1}{{\sqrt N }}\sum\limits_{i = 1}^{N} {\left| i\right\rangle \left| {{a_i}} \right\rangle \left| {{b_i}} \right\rangle  \otimes \left| 0 \right\rangle }$.}
\label{Fig. 3 }
\end{figure}
\begin{equation}\label{5}
\frac{1}{{\sqrt N }}\sum\limits_{i = 1}^{N} {\left| i\right\rangle \left| {{a_i}} \right\rangle \left| {{b_i}} \right\rangle  \otimes \left| 0 \right\rangle } \xrightarrow{{{O_f}}}\frac{1}{{\sqrt N }}\sum\limits_{i = 1}^{N} {\left| i\right\rangle \left| {{a_i}} \right\rangle \left| {{b_i}} \right\rangle \left| {{\kern 1pt} f({a_i},{b_i})} \right\rangle } {\kern 1pt} ,
\end{equation}
\begin{equation}\label{6}
\frac{1}{{\sqrt N }}\sum\limits_{i = 1}^{N} {\left| i\right\rangle \left| {{b_i}} \right\rangle \left| {{a_i}} \right\rangle \otimes \left| 0 \right\rangle } \xrightarrow{{{O_f}}}\frac{1}{{\sqrt N }}\sum\limits_{i = 1}^{N} {\left| i\right\rangle \left| {{b_i}} \right\rangle \left| {{a_i}} \right\rangle \left| {{\kern 1pt} f({a_i},{b_i})} \right\rangle } {\kern 1pt} .
\end{equation}
Then Bob (Alice) gets the final state  $\left| {{{\phi '''}_A}} \right\rangle  = \frac{1}{{\sqrt N }}\sum\limits_{i = 1}^{N} {\left| i\right\rangle \left| {{a_i}} \right\rangle \left| {{b_i}} \right\rangle \left| {f({a_i},{b_i})} \right\rangle }$ ($\left| {{{\phi '''}_B}} \right\rangle  = \frac{1}{{\sqrt N }}\sum\limits_{i = 1}^{N} {\left| i\right\rangle \left| {{a_i}} \right\rangle \left| {{b_i}} \right\rangle \left| {f({a_i},{b_i})} \right\rangle }$). Here, the function $f$ is a qubit comparator defined as below,
\begin{equation}\label{7}
f({a_i},{b_i}) = \left\{ 	
\begin{gathered}
  1,{\kern 1pt} {\kern 1pt} {\kern 1pt} {\kern 1pt} {a_i} \ge {b_i} \hfill \\
  0,{\kern 1pt} {\kern 1pt} {\kern 1pt} {a_i} < {b_i}{\kern 1pt}  \hfill \\
\end{gathered}  \right..
\end{equation}
And Fig. 4 and Fig. 5 give the corresponding circuit implementation of the qubit comparator.

\begin{figure}[htb]
\centering
 \includegraphics[width=2.8in]{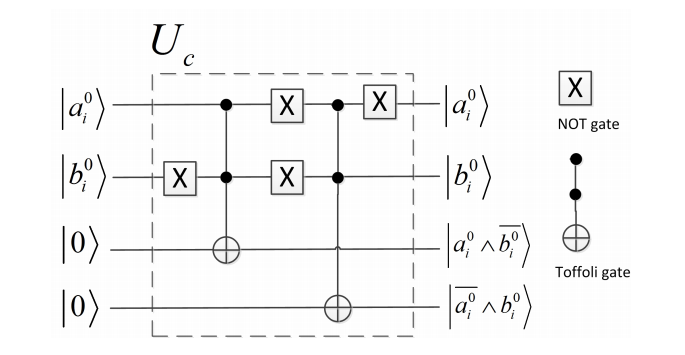}
\DeclareGraphicsExtensions.
\caption{The circuit of one-qubit comparator comparing two qubits. Here, $\times $ is the NOT gate, and $\bullet$ in the Toffloi gate denotes the negative-control qubit conditional being set to one, while $\oplus$ represents the target qubit.}
\label{Fig. 4 }
\end{figure}

\begin{figure}[htb]
\centering
 \includegraphics[width=3.5in]{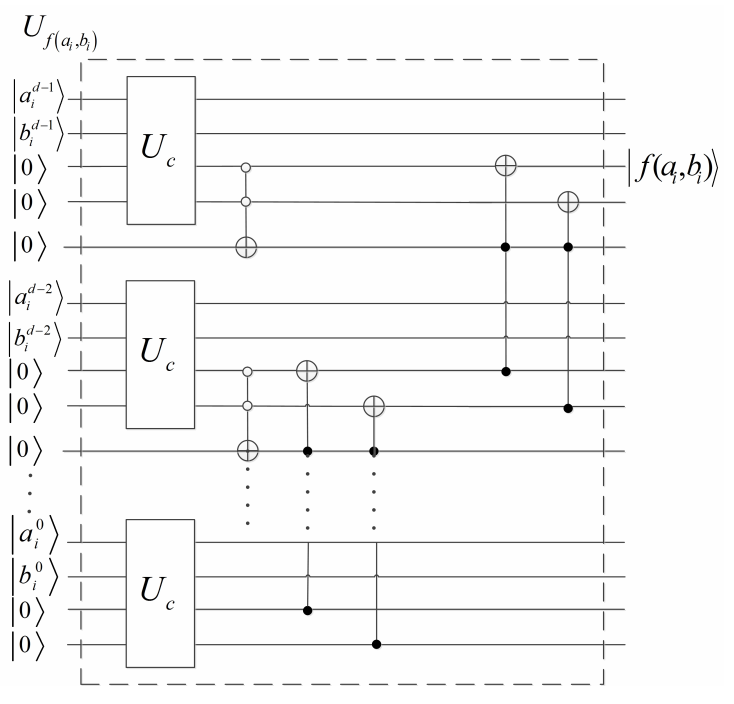}
\DeclareGraphicsExtensions.
 \caption{The circuit of multi-qubit comparator comparing ${a_i}$ and ${b_i}$. ${U_c}$ is the one-qubit comparator described in Fig. 3, and $\circ$ denotes the positive-control qubit conditional being set to zero. }
\label{Fig. 5 }
\end{figure}

Step 4. Bob (Alice) executes the quantum counting algorithm~\cite{BQ98} on $\left| {{{\phi '''}_A}} \right\rangle$ ($\left| {{{\phi '''}_B}} \right\rangle$)  to obtain the number of those products which meet $f({a_i},{b_i}) = 1$, respectively. Then he (she) will get the estimator of  counting result ${t_B} = \sum\limits_{i = 1}^{N} {f\left( {{a_i},{b_i}} \right)}$ (${t_A} = \sum\limits_{i = 1}^{N} {f\left( {{a_i},{b_i}} \right)}$).

Step 5. Alice and Bob execute quantum bit string commitment~\cite{KQ03} to exchange $n$-bit string ${t_A}$ and ${t_B}$,
(1) Alice (Bob) commits ${t_A}$ (${t_B}$) by preparing the $logm$-qubit state ${\left| \tau  \right\rangle _A}$ (${\left| \tau  \right\rangle _B}$) according to an error correcting code~\cite{BQ01} and sending them to Bob(Alice), here $m=cn$ for a fixed constant $c > 1$;
(2) Alice(Bob) unveils ${t_A}$ (${t_B}$) to the other.
(3) Alice (Bob) verifies the validness of the other's bit string by measuring the projection onto ${\left| \tau  \right\rangle _B}$ (${\left| \tau  \right\rangle _A}$). If he(she) obtains eigenvalue 1, he (she) accepts the unveiling; otherwise he or she concludes Bob(Alice) cheated.
Then, Alice and Bob check the other's honesty by comparing the two counting results, if both of the two parties are honest, the difference of ${t_A}$ and ${t_B}$ will be less than $\delta$ (i.e., $\left| {{t_A} - {t_B}} \right| \le \delta$), where $\delta$ is total error of estimation in quantum counting algorithm.

Step 6.  Alice and Bob further check whether the two counting results satisfy the trading requirement: ${t_A} \ge \varepsilon$ and ${t_B} \ge \varepsilon$ ($1 \ll \varepsilon \le N$). If they meet the requirement, Alice and Bob continue to trade, otherwise they will cancel this negotiation.

\section{Correctness, Privacy and Efficiency Analysis}
\label{sec:4}
\subsection{Correctness analysis}
\label{sec:4.1}
\textbf{Theorem 1.} \emph{Assume there exist a constant $\varepsilon$, $1 \ll \varepsilon  \le N$, such that for two participants Alice and Bob executing our protocol can satisfy the correctness property of Q3PEN protocol defined in Sect. 2, that is, two honest parties Alice and Bob can get the correct result of whether  $\sum\limits_{i = 1}^N {f({a_i},{b_i})}  \ge \varepsilon $.}
\\\\
\noindent\emph{Proof:} Without loss of generality, we suppose the number of products $N=6$, the threshold $\varepsilon=5$, and Alice's maximal buying prices  $A = (3, 2, 5, 4, 7, 6)$, Bob's minimal selling prices $B=(2, 2, 5, 5, 6, 6)$, so $n = \left\lceil {\log (6 + 1)} \right\rceil=3$, $d = \left\lceil {\log \left( {\mathop {\max }\limits_{i,{\kern 1pt} {\kern 1pt} k} ({a_j},{b_k}) + 1} \right)} \right\rceil=\left\lceil {\log (7 + 1)} \right\rceil=3$. In Step 1, Alice and Bob prepare states $\left| {{{\phi '}_A}} \right\rangle$ and $\left| {{{\phi '}_B}} \right\rangle$ through the oracle operations ${O_A}$ and ${O_B}$,
\begin{equation}\label{8}
\left| {{{\phi '}_A}} \right\rangle  = \frac{1}{{\sqrt 6 }}(\left| 1 \right\rangle \left| 3 \right\rangle  + \left| 2 \right\rangle \left| 2 \right\rangle  + \left| 3 \right\rangle \left| 5 \right\rangle  + \left| 4 \right\rangle \left| 4 \right\rangle  + \left| 5 \right\rangle \left| 7 \right\rangle  + \left| 6 \right\rangle \left| 6 \right\rangle ),
\end{equation}
\begin{equation}\label{9}
\left| {{{\phi '}_B}} \right\rangle  = \frac{1}{{\sqrt 6 }}(\left| 1 \right\rangle \left| 2 \right\rangle  + \left| 2 \right\rangle \left| 2 \right\rangle  + \left| 3 \right\rangle \left| 5 \right\rangle  + \left| 4 \right\rangle \left| 5 \right\rangle  + \left| 5 \right\rangle \left| 6 \right\rangle  + \left| 6 \right\rangle \left| 6 \right\rangle ),
\end{equation}
and then send them to each other, respectively. After receiving $\left| {{{\phi '}_A}} \right\rangle$ and $\left| {{{\phi '}_B}} \right\rangle$ in Step 2, Bob(Alice) applies the oracle operation  ${O_B}$(${O_A}$) again, and they get the new state as below,
\begin{equation}\label{10}
\begin{gathered}
  \left| {{{\phi ''}_A}} \right\rangle  = \frac{1}{{\sqrt 6 }}(\left| 1 \right\rangle \left| 3 \right\rangle \left| 2 \right\rangle  + \left| 2 \right\rangle \left| 2 \right\rangle \left| 2 \right\rangle  + \left| 3 \right\rangle \left| 5 \right\rangle \left| 5 \right\rangle {\kern 1pt} {\kern 1pt}  + \left| 4 \right\rangle \left| 4 \right\rangle \left| 5 \right\rangle {\kern 1pt}  \hfill \\
  {\kern 1pt} {\kern 1pt} {\kern 1pt} {\kern 1pt} {\kern 1pt} {\kern 1pt} {\kern 1pt} {\kern 1pt} {\kern 1pt} {\kern 1pt} {\kern 1pt} {\kern 1pt} {\kern 1pt} {\kern 1pt} {\kern 1pt} {\kern 1pt} {\kern 1pt} {\kern 1pt} {\kern 1pt} {\kern 1pt} {\kern 1pt} {\kern 1pt} {\kern 1pt} {\kern 1pt} {\kern 1pt} {\kern 1pt} {\kern 1pt} {\kern 1pt} {\kern 1pt} {\kern 1pt} {\kern 1pt}  + \left| 5 \right\rangle \left| 7 \right\rangle \left| 6 \right\rangle {\kern 1pt}  + \left| 6 \right\rangle \left| 6 \right\rangle \left| 6 \right\rangle ). \hfill \\
\end{gathered}
\end{equation}
\begin{equation}\label{11}
\begin{gathered}
  \left| {{{\phi ''}_B}} \right\rangle  = \frac{1}{{\sqrt 6 }}(\left| 1 \right\rangle \left| 2 \right\rangle \left| 3 \right\rangle  + \left| 2\right\rangle \left| 2 \right\rangle \left| 2 \right\rangle  + \left| 3 \right\rangle \left| 5 \right\rangle \left| 5 \right\rangle  + \left| 4 \right\rangle \left| 5 \right\rangle \left| 4 \right\rangle  \hfill \\
  {\kern 1pt} {\kern 1pt} {\kern 1pt} {\kern 1pt} {\kern 1pt} {\kern 1pt} {\kern 1pt} {\kern 1pt} {\kern 1pt} {\kern 1pt} {\kern 1pt} {\kern 1pt} {\kern 1pt} {\kern 1pt} {\kern 1pt} {\kern 1pt} {\kern 1pt} {\kern 1pt} {\kern 1pt} {\kern 1pt} {\kern 1pt} {\kern 1pt} {\kern 1pt} {\kern 1pt} {\kern 1pt} {\kern 1pt} {\kern 1pt} {\kern 1pt} {\kern 1pt} {\kern 1pt} {\kern 1pt}  + \left| 5 \right\rangle \left| 6 \right\rangle \left| 7 \right\rangle {\kern 1pt}  + \left| 6\right\rangle \left| 6 \right\rangle \left| 6 \right\rangle ). \hfill \\
\end{gathered}
\end{equation}
Bob (Alice) further applies another oracle operation ${O_f}$ on $\left| {{{\phi ''}_A}} \right\rangle$ ($\left| {{{\phi ''}_B}} \right\rangle$), and embed the comparative results of buying and selling prices into the qubit in the last position,
\begin{equation}\label{12}
\begin{gathered}
  \left| {{{\phi '''}_A}} \right\rangle  = \frac{1}{{\sqrt 6 }}(\left| 1 \right\rangle \left| 3 \right\rangle \left| 2 \right\rangle \left| 1 \right\rangle  + \left| 2 \right\rangle \left| 2 \right\rangle \left| 2 \right\rangle \left| 1 \right\rangle  + \left| 3 \right\rangle \left| 5 \right\rangle \left| 5 \right\rangle \left| 1 \right\rangle  \hfill \\
  {\kern 1pt} {\kern 1pt} {\kern 1pt} {\kern 1pt} {\kern 1pt} {\kern 1pt} {\kern 1pt} {\kern 1pt} {\kern 1pt} {\kern 1pt} {\kern 1pt} {\kern 1pt} {\kern 1pt} {\kern 1pt} {\kern 1pt} {\kern 1pt} {\kern 1pt} {\kern 1pt} {\kern 1pt} {\kern 1pt} {\kern 1pt} {\kern 1pt} {\kern 1pt} {\kern 1pt} {\kern 1pt} {\kern 1pt} {\kern 1pt} {\kern 1pt} {\kern 1pt}  + \left| 4 \right\rangle \left| 4 \right\rangle \left| 5 \right\rangle \left| 0 \right\rangle {\kern 1pt}  + \left| 5 \right\rangle \left| 7 \right\rangle \left| 6 \right\rangle {\kern 1pt} \left| 1 \right\rangle  + \left| 6 \right\rangle \left| 6 \right\rangle \left| 6 \right\rangle \left| 1 \right\rangle ). \hfill \\
\end{gathered}
\end{equation}
\begin{equation}\label{13}
\begin{gathered}
  \left| {{{\phi '''}_B}} \right\rangle  = \frac{1}{{\sqrt 6 }}(\left| 1 \right\rangle \left| 2 \right\rangle \left| 3 \right\rangle \left| 1 \right\rangle  + \left| 2 \right\rangle \left| 2 \right\rangle \left| 2 \right\rangle \left| 1 \right\rangle  + \left| 3 \right\rangle \left| 5 \right\rangle \left| 5 \right\rangle \left| 1 \right\rangle  \hfill \\
  {\kern 1pt} {\kern 1pt} {\kern 1pt} {\kern 1pt} {\kern 1pt} {\kern 1pt} {\kern 1pt} {\kern 1pt} {\kern 1pt} {\kern 1pt} {\kern 1pt} {\kern 1pt} {\kern 1pt} {\kern 1pt} {\kern 1pt} {\kern 1pt} {\kern 1pt} {\kern 1pt} {\kern 1pt} {\kern 1pt} {\kern 1pt} {\kern 1pt} {\kern 1pt} {\kern 1pt} {\kern 1pt} {\kern 1pt} {\kern 1pt} {\kern 1pt} {\kern 1pt}  + \left| 4 \right\rangle \left| 5 \right\rangle \left| 4 \right\rangle \left| 0 \right\rangle {\kern 1pt}  + \left| 5 \right\rangle \left| 6 \right\rangle \left| 7 \right\rangle {\kern 1pt} \left| 1 \right\rangle  + \left| 6 \right\rangle \left| 6 \right\rangle \left| 6 \right\rangle \left| 1 \right\rangle ). \hfill \\
\end{gathered}
\end{equation}
In Step 4, Bob and Alice further execute quantum counting to get their counting result ${t_A} = \sum\limits_{i = 1}^6 {f\left( {{a_i},{b_i}} \right)}  = 5$, ${t_B} = \sum\limits_{i = 1}^6 {f\left( {{a_i},{b_i}} \right)}  = 5$, respectively. Thus, the result value ${t_A} = {t_B} \ge 5$, so the correctness of the protocol is guaranteed.

\subsection{Privacy analysis}
\label{sec:4.2}

The privacy property means that both Alice and Bob cannot get the other's price information or any secret information deduced from the final result, which consists of Bob's or Alice's privacy.

\subsubsection{Alice's privacy}

\textbf{Theorem 2.} \emph{Assume $N$ is the number of products, $n = \left\lceil {\log (N + 1)} \right\rceil$, and $m$ is the number of commited bits ($m=cn$, $c>1$), such that Bob cannot learn any secret information about Alice's prices $A$, the possible information deduced from $\sum\limits_{i = 1}^N {f({a_i},{b_i})}$ and his prices $B$. And Bob's cheating behavior can be detected by Alice with probability ($1 - {\left( {\frac{1}{2}} \right)^{m - \log n}}$) very close to 1.}
\\\\
\noindent\emph{Proof:} Suppose Bob is dishonest and he tries to take some strategies to get information about Alice's private prices $A$.  In our protocol,  all the classical information about $A$ is embedded into the state  $\left| {{{\phi '}_A}} \right\rangle  = \frac{1}{{\sqrt N }}\sum\limits_{i = 1}^{N} {\left| i\right\rangle \left| {{a_i}} \right\rangle }$,  so the most possible attack is that Bob performs a certain projective measurement on state $\left| {{{\phi '}_A}} \right\rangle$ when he received it from Alice in Step 2. Then, Bob will get $\left| i \right\rangle \left| {{a_i}} \right\rangle $  with the same probability $\frac{1}{N}$ for any $i$. Thus $\left| {{\phi '}_A} \right\rangle $ can be characterized by the quantum ensemble $\varepsilon  = \{ {P_i},{\rho}(i)\} $, where ${P_i} = \frac{1}{N}$ is the probability that Bob gets the component ${a_i}$ (Assuming that Bob does not have any prior information on Alice's prices initially), and
\begin{equation}\label{14}
{\rho}(i) = \left| i \right\rangle \left| {{a_i}} \right\rangle \left\langle {{a_i}} \right|\left\langle i \right|.
\end{equation}
According to Holevo's theorem~\cite{HP82}, the information that Bob can access by performing any measurement on $\left| {{\phi '}_A} \right\rangle $ is bounded by the entropy:
\begin{equation}\label{15}
\begin{gathered}
  I \le X(\varepsilon ) = S({\rho}) - \frac{1}{N}\sum\limits_{i = 1}^N {S({\rho}(i))}  \hfill \\
  {\kern 1pt} {\kern 1pt} {\kern 1pt} {\kern 1pt} {\kern 1pt} {\kern 1pt} {\kern 1pt} {\kern 1pt} {\kern 1pt} {\kern 1pt} {\kern 1pt} {\kern 1pt} {\kern 1pt} {\kern 1pt} {\kern 1pt} {\kern 1pt} {\kern 1pt} {\kern 1pt} {\kern 1pt} {\kern 1pt} {\kern 1pt} {\kern 1pt} {\kern 1pt} {\kern 1pt} {\kern 1pt} {\kern 1pt} {\kern 1pt} {\kern 1pt} {\kern 1pt} {\kern 1pt} {\kern 1pt} {\kern 1pt} {\kern 1pt} {\kern 1pt} {\kern 1pt} {\kern 1pt} {\kern 1pt} {\kern 1pt} {\kern 1pt} {\kern 1pt} {\kern 1pt} {\kern 1pt} {\kern 1pt} {\kern 1pt} {\kern 1pt} {\kern 1pt}  = S({\rho}) \hfill \\
\end{gathered}
\end{equation}
where ${\rho} = \sum\limits_{i = 1}^N {\left( {\frac{1}{N}{\rho}(i)} \right)}$, so $I \le S({\rho}) = \log N$. From the above analysis, we can see that Bob can obtain at most one element of Alice's prices $A$, i.e., one-$N$th of Alice's private information, by any possible measurement. And he will loses the chance to get the other elements at the same time. What is more, the information that Bob can extract can be ignored when number $N$ is large enough.

Besides, if Bob has performed the projective measurement to extract some information about $A$, he will certainly lose the chance to further get his counting result  ${t_B}$. However, in order to complete the honest test of Alice in Step 5, Bob may try to steal
as much information as possible from Alice's quantum commitment. Bob can extract at most $\log n$ bits from Alice's $m$-length quantum bit string ($m=cn$, $c>1$)~\cite{KQ03}. Furthermore, the probability that Bob successfully stealing all $n$ bits is at most $P = {\left( {\frac{1}{2}} \right)^{m - \log n}}$. Therefore, if Bob extracts Alice's partial private information from the state $\left| {{{\phi '}_A}} \right\rangle$, his cheating behavior will be detected by Alice's honest test with at least the probability of ($1 - P$), and the probability will be very close to $1$ when $n$ is large enough.

\subsubsection{Bob's privacy}

\textbf{Theorem 3.} \emph{Assume $N$ is the number of products, $n = \left\lceil {\log (N + 1)} \right\rceil$, and $m$ is the number of commited bits ($m=cn$, $c>1$), such that Alice cannot learn any secret information about Bob's prices $B$, the possible information deduced from $\sum\limits_{i = 1}^N {f({a_i},{b_i})}$ and his prices $A$. And Alice's cheating behavior can be detected by Bob with probability ($1 - {\left( {\frac{1}{2}} \right)^{m - \log n}}$) very close to 1.}
\\\\
\noindent\emph{Proof:} Similarly, a dishonest Alice can obtain at most one element of Bob's private prices $B$ (i.e., ${b_i}$) by her local measurement on state $\left| {{{\phi '}_B}} \right\rangle$. However, she will certainly lose the chance to further get the counting result  ${t_B}$, and her cheating behavior will be detected by Bob with the probability of ($1 - {\left( {\frac{1}{2}} \right)^{m - \log n}}$).

\subsection{Efficiency analysis}
The communication cost is one of the key indicators of the efficiency for communication protocols. In our protocol, the buyer Alice and the seller Bob exchange two quantum state, $\left| {{\psi _A}} \right\rangle $ and  $\left| {{\psi _B}} \right\rangle $, with each other in Step 1, where each quantum state is the $(n+d)$-qubit state. Besides, the counting results $t_A$ and $t_B$ are exchanged too in Step 5, where $t_A$ and $t_B$ all are the classic $n$-bit message. Therefore, the total communication cost of our protocol is $2(n + d)$ qubits and $2n$ cbits,  where $n = \left\lceil {\log (N + 1)} \right\rceil$, $d = \left\lceil {\log \left( {\mathop {\max }\limits_{i,{\kern 1pt} {\kern 1pt} k} ({a_j},{b_k}) + 1} \right)} \right\rceil$ (i.e., $d$ is just relevant to the range of selling and buying prices and there is no direct correlation between $d$ and $N$). So, the communication complexity of our protocol is $O\left( {logN} \right)$. However, in the classic C05 protocols~\cite{CP05}, $2N$ cbits are exchanged between the buyer and the seller, while $4N$ cbits is required to be transmitted in the A07 protocol~\cite{AS07} (it should be noted that the message of product's price is just one cbit in  these classic protocols),  which means both of C05 and A07 protocols hold the linear communication complexity $O\left( {N} \right)$. Obviously, our protocol achieves a great reduction in the communication complexity.

In order to more intuitively demonstrate the difference of communication costs between our Q3PEN protocol and  the classic C05, A07 protocols,  we uniformly suppose the message of each product's price is two cbits,  i.e., $d = 2$.  So  the exchanged message of the C05 and A07 protocols are $2N*d=4N$ and $4N*d=8N$ cbits, respectively, while the exchanged qubits/cbits  in our protocol are $4\left\lceil{\log(N+1)}\right\rceil+2d=4\left\lceil {\log (N + 1)} \right\rceil+4$. And Fig. 6 illustrates the communication costs among the classic C05, A07 protocols and our Q3PEN protocol.

\begin{figure}[htb]
\centering
 \includegraphics[width=3.5in]{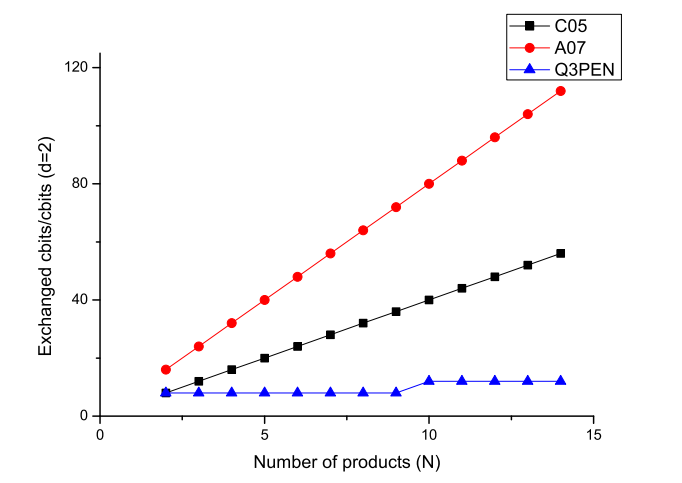}
\DeclareGraphicsExtensions.
 \caption{The communication costs of the C05, A07 protocols and our Q3PEN protocol when $d=2$.}
\label{Fig. 6 }
\end{figure}

\section{Conclusion}
With ingeniously making  use of quantum oracle and quantum counting technology, we propose an efficient quantum privacy-preserving price e-negotiation protocol. Compared with  those previous classic protocols, the communication costs is greatly reduced, and the two participants' privacies are perfectly guaranteed as well. Although the proposed protocol is an example of two-party price e-negotiation, it can be extended to the multi-party scenario, such as one-to-many (i.e., one buyer negotiates with $n$ sellers), many-to-one, and many-to-many.

\begin{acknowledgements}
The authors would like to thank the anonymous reviewers and editor for their comments that improved the quality of this paper. This work is supported by National Natural Science Foundation of China (Grant Nos. 61672290 and 61802002), Natural Science Foundation of Jiangsu Province(Grant No. BK20171458), the Six Talent Peaks Project of Jiangsu Province (Grant No. 2015-XXRJ-013), the Practice Innovation Training Program Projects for Jiangsu College Students (Grant No. 201810300016Z), and the Priority Academic Program Development of Jiangsu Higher Education Institutions(PAPD).
\end{acknowledgements}



\end{document}